\documentclass[prl,twocolumn,superscriptaddress]{revtex4-1}

\usepackage[dvipdfmx]{graphicx}
\usepackage{bmpsize}
\usepackage{dcolumn}
\usepackage{bm}
\usepackage{color}

\graphicspath{{fig/}}

\begin{document}

\preprint{APS/123-QED}

\title{
Hidden robust presence of a hole Fermi surface in a  
heavily electron doped iron based superconductor LaFe$_2$As$_2$
} 

\author{Hidetomo Usui}
\affiliation{Department of Physics and Materials Science, Shimane University, 1060 Nishikawatsu-cho, Matsue, Shimane, 690-8504, Japan}
\author{Kazuhiko Kuroki}
\affiliation{Department of Physics, Osaka University, 1-1 Machikaneyama-cho, Toyonaka, Osaka, 560-0043, Japan}

\date{\today}

\begin{abstract}
We investigate the electronic structure of a recently discovered, 
heavily electron-doped iron based superconductor LaFe$_2$As$_2$. 
Although first principles calculation shows apparent absence of 
hole Fermi surfaces 
around the $\Gamma$ point, we reveal, by hypothetically removing 
the La $d$ orbital contribution, that a hole Fermi surface around 
the $\Gamma$ point is essentially present.
In the collapsed phase of LaFe$_2$As$_2$, which is non-superconducting, 
the hole Fermi surface is found to be absent, and the difference from 
the uncollapsed superconducting phase can be naturally understood within 
the spin-fluctuation mediated pairing scenario.
\end{abstract}

\pacs{ }
\maketitle

From the early stage of the study, the Fermi surface configuration in  
the iron based superconductors\cite{Kamihara2008} has been an issue of great interest. This is because the Fermi surface consists of 
small electron and hole pockets (or cylinders in the three dimensional sense), 
so that it is sensitive against carrier doping, elemental substitution\cite{Mazin2008,Kuroki2009PRB}, 
and/or pressurization, and also because many theories suggest that the number and/or asymmetry of 
the electron and hole Fermi surfaces play a key role in controlling 
the superconducting transition temperature ($T_c$) and/or the pairing symmetry
\cite{Hirschfeld2011review,Hosono2015review}.
For instance, within the spin-fluctuation-mediated pairing scenario, 
Cooper pair scattering occurs between electron and hole Fermi surfaces, 
so that too much carrier doping is expected to result in a disappearance of 
hole or electron Fermi surfaces, and hence suppression or absence of superconductivity. 
$T_c$ dome obtained in a case when, e.g., electrons are introduced in BaFe$_2$As$_2$ 
by partially substituting Fe with Co\cite{Sefat2008,Chu2009} is considered to be
 a manifestation of such a Fermi surface variance.\cite{Fang2009}

From this viewpoint, cases with heavy electron doping have attracted much attention. 
In K$_x$Fe$_{2-y}$Se$_2$, a large amount of electrons are doped, and in fact, 
the angle resolved photoemission spectroscopy (ARPES) experiments show that 
the hole bands sink below the Fermi level\cite{Guo2010_ironbased122, Qian2011_ironbased122FS, Niu2015_ironbased11}. 
Still, $T_c$ is high, apparently implying that the hole bands are not playing an important role in the occurrence of superconductivity. 
A similar situation (with even a higher $T_c$) is observed in FeSe thin films grown on 
a substrate\cite{Wang2012_ironbased11-STO, Tan2013_ironbased11-STO,Nojima}. 
Partially motivated by these studies, the importance of the hole bands 
sinking below the Fermi level (the ``incipient band'') has been pointed out in 
various studies\cite{Hirschfeld2011_incipient, Miao2015_ironbased111FS, Wang2011_122FRG, 
Bang2014_122shadowgap, Chen2015_incipient, Bang2016_dynamicaltuning,Kuroki2005_wide-narrow,
Matsumoto2018,Aoki2016,MaierScalapino2019}. 
On the other hand, another ARPES experiment on K$_x$Fe$_{2-y}$Se$_2$ found that a hole band actually intersects the Fermi level\cite{Sunagawa}. 
Another series of material that is of great interest are the hydrogen-doped 1111 compounds 
such as LaFeAsO$_{1-x}$H$_x$.\cite{Iyo2010H,Hosono2011,Iimura2012,Hiraishi2014,Iimura2017,Matsuishi2014}  
In these materials, high $T_c$ superconductivity survives up to large amount of electron doping close to $50\%$. 
Although such heavy electron doping is expected to wipe out the hole Fermi surface, 
some theoretical studies have pointed out that one of the hole bands (the $d_{xy}$ band) actually 
exhibits a non-rigid-band shift upon electron (hydrogen) doping, 
so that the hole band can in fact give rise to a Fermi surface and play an important role\cite{Suzuki2014,Iimura2017}.

Given this controversial situation regarding heavily electron-doped iron-based superconductors, 
a recent discovery of a new superconductor LaFe$_2$As$_2$, exhibiting $T_c=12.1$ K, 
opens up a renewed avenue in this hotly debated field.\cite{Iyo2019} 
In fact, considering BaFe$_2$As$_2$ as the mother (starting) compound, 
a complete substitution of Ba$(+2)$ with La$(+3)$ corresponds to a very heavy electron doping of $50\%$. 
A band structure calculation of LaFe$_2$As$_2$ have shown that 
the hole Fermi surface around the $\Gamma$ point is indeed missing\cite{Iyo2019}, which may be taken as natural considering the large amount of electron doping. 
This study also found that one electron Fermi surface interestingly possesses 
an unusual ``jungle-gym-like" shape, whose role played in superconductivity, if any, is difficult to understand.
It has also been found that the crystal structure transforms from 
a collapsed tetragonal structure to an uncollapsed tetragonal structure by annealing samples.\cite{Iyo2019}
The superconductivity is found to emerge only in the uncollapsed phase, 
and no superconductivity has been observed in the collapsed phase.
As discussed for other iron based superconductors\cite{Hirschfeld2011review, Dhaka2014}, it is expected that 
the difference in the the electronic structure of collapsed and uncollapsed 
LaFe$_2$As$_2$ may hold the key for understanding the origin of superconductivity.

Here, we theoretically investigate the electronic structure of 
the uncollapsed and collapsed phases of LaFe$_2$As$_2$ in an aim to 
reveal its relevance to the origin of superconductivity. 
In the uncollapsed phase, we surprisingly find that the $d_{xy}$ cylindrical hole Fermi surface 
around the $\Gamma$ point, in the usual sense of the term used for the iron-based superconductors, 
can be considered as essentially present despite the heavy electron doping. 
Interestingly, both the survival of the $d_{xy}$ hole Fermi surface and its apparent absence can be traced back to the hybridization between the Fe $3d$ and La $5d$ orbitals.
In the collapsed phase, on the other hand, the $d_{xy}$ hole Fermi surface is lost 
due to the large Fe-As-Fe bond angle. The correspondence between the presence/absence of 
the hole Fermi surface and the $T_c$ can be naturally understood within the spin-fluctuation-mediated pairing scenario.

First principles calculations are performed by means of the full-potential linearized augmented plane wave method 
as implemented in WIEN2k\cite{Wien2k}.
We use the PBE exchange-correlation functional\cite{PBE}, and take 
$RK_{\rm max} = 7$ and a $16 \times 16 \times 7$ $k$-mesh for self-consistent calculation
and a $22 \times 22 \times 22$ $k$-mesh for calculation of the density of states.
We start with the electronic structure of the uncollapsed phase, whose 
lattice constants and the internal coordinates of the Fe atoms are given in Ref. \cite{Iyo2019}. 

\begin{figure}
	\includegraphics[width=7cm]{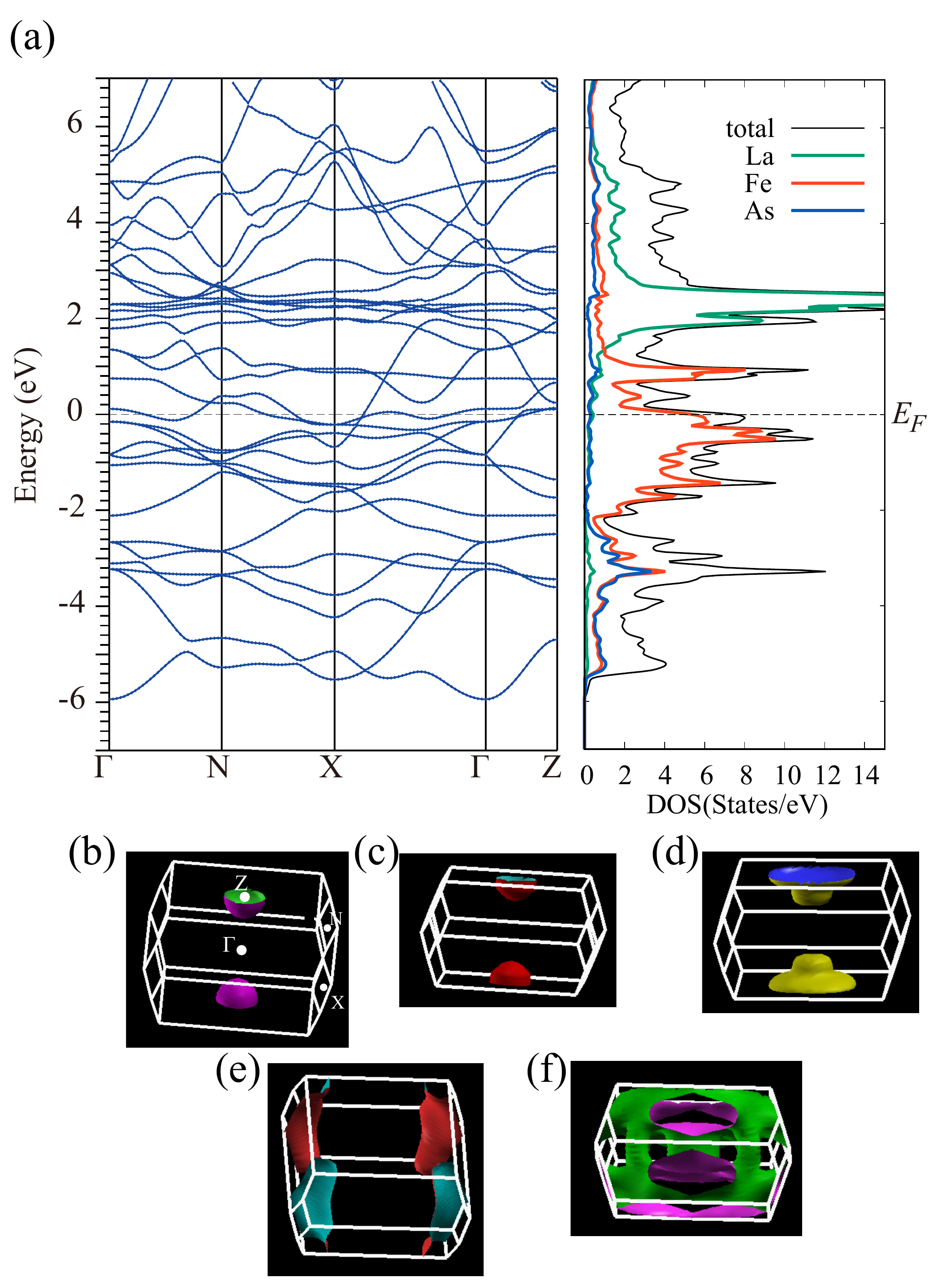}
	\caption{(a) The electronic band structure and the density of states of uncollapsed LaFe$_2$As$_2$.
	(b)-(f) The five Fermi surfaces of LaFe$_2$As$_2$.}
	\label{fig1}
\end{figure}

The calculated band structure, the density of states and the Fermi surface of the uncollapsed phase are 
shown in Fig. \ref{fig1}. 
In Fig. \ref{fig1}(a), it can be seen that the Fe $d$ orbitals are 
dominant for the construction of the conduction bands around the Fermi level.
As found in Ref. \cite{Iyo2019}, the three hole and two electron Fermi surfaces are obtained as shown in Figs. \ref{fig1}(b)-(d) and (e)-(f), respectively.
The hole Fermi pockets and the cylindrical electron Fermi surface shown in Figs. \ref{fig1}(a)-(e) 
are obtained, which can be seen in other iron based superconductors.
On the other hand, the electron Fermi surface shown in Fig. \ref{fig1}(f) does not seem to 
correspond to any of the Fermi surfaces seen in usual iron based superconductors.

\begin{figure}
	\includegraphics[width=7cm]{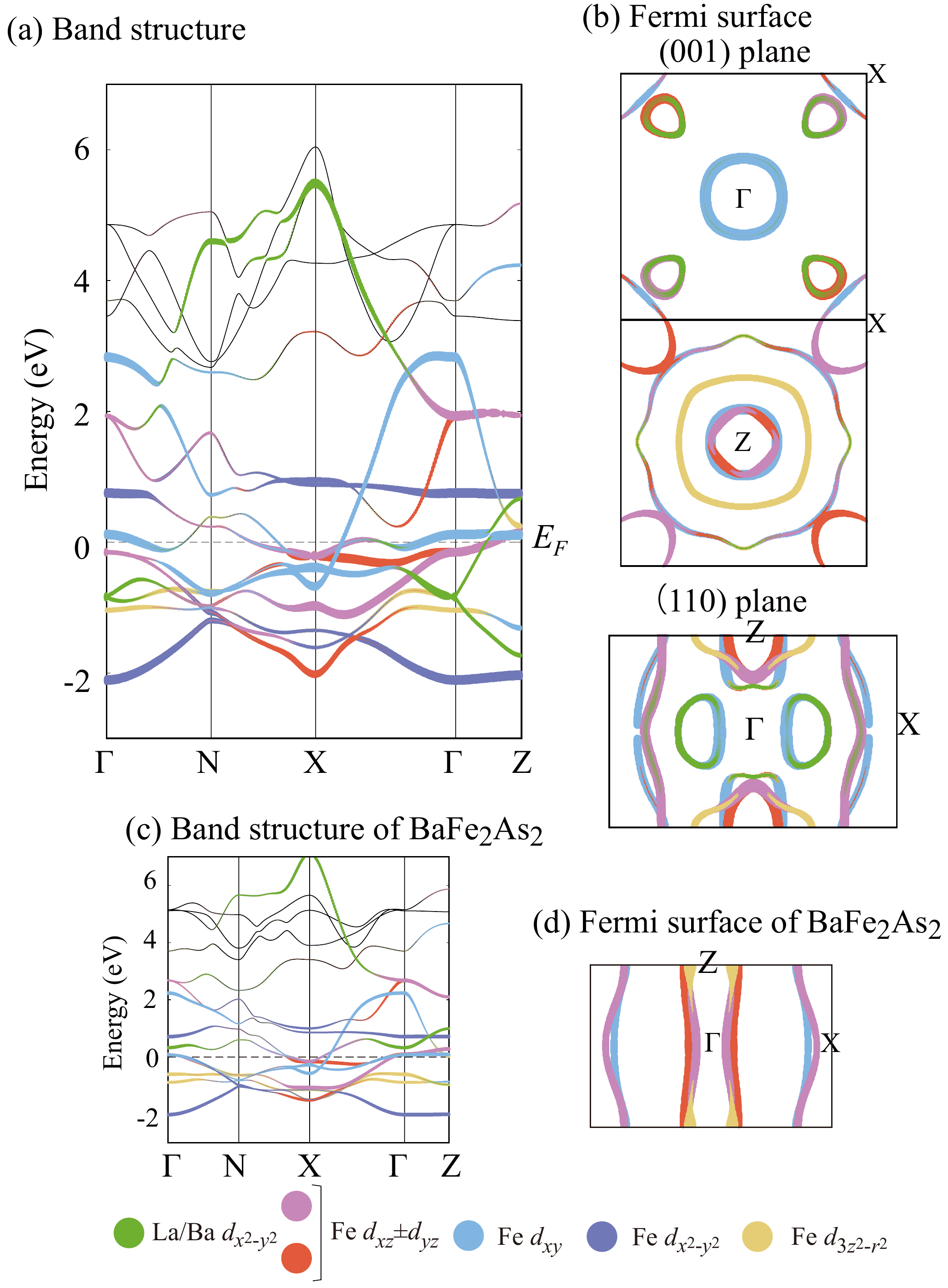}
	\caption{(a) The band structure and (b) the Fermi surface  of uncollapsed LaFe$_2$As$_2$ obtained from the 15-orbital tight binding model.
	(c) and (d) are the band structure and the Fermi surface of paramagnetic BaFe$_2$As$_2$, respectively, obtained in a similar way.}
	\label{fig2}
\end{figure}

In order to understand the origin of this strange Fermi surface, 
we construct a 15-orbital tight binding model for LaFe$_2$As$_2$ exploiting the 
maximally localized Wannier functions of $d$ orbital projections centered at 
the Fe or La sites (5 $d$ orbitals $\times$ 3 atoms = 15 orbitals) using wannier90 package\cite{Wannier} and wien2wannier code\cite{w2w}. 
The band structure and the Fermi surface of the tight binding model  
are shown in Figs. \ref{fig2} (a) and (b), respectively.
In these figures, the orbital component of the La $d_{x^2-y^2}$ and Fe $d$ orbitals are expressed as colored circles. Other La $d$ orbitals have only small weight within the energy window presented here. 
For comparison, as shown in Figs. \ref{fig2} (c) and (d),
we also perform similar calculation for BaFe$_2$As$_2$ (assuming a paramagnetic state) 
adopting the experimentally determined lattice structure\cite{Rotter2008}.

First glance at the density of states of LaFe$_2$As$_2$ (Fig. \ref{fig1}(a)) might give 
an impression that the La orbitals do not strongly affect the band structure around the Fermi level. 
However, the La $d_{x^2-y^2}$ orbital weight is seen to be spread from $-2$ to 6 eV measured from the Fermi level, 
namely, the La $d_{x^2-y^2}$ band crosses the Fermi level. 
In fact, if we look at the orbital weight in the band structure and the Fermi surface, 
we find that the La $d_{x^2-y^2}$ orbital does have large contribution (see the green colored circles in Figs. \ref{fig2}(a) and (b)). 
The ``jungle-gym-like" shape of the Fermi surface therefore can be understood as 
originating from the hybridization between the Fe 3$d$ and La $d_{x^2-y^2}$ orbitals. 
This is why the strange shape is hardly seen in other iron based superconductors. 
In fact, in BaFe$_2$As$_2$, nearly the entire Ba $d_{x^2-y^2}$ orbital weight lies 
above the Fermi level (Fig. \ref{fig2}(c)), so that the Ba $d_{x^2-y^2}$ orbital has small contribution on 
the Fermi surface.   
We note that the presence of La $d$ orbital component around the Fermi level has also been discussed in LaFe$_2$P$_2$, 
which belongs to the same space group.\cite{Morsen1988, Razzoli2015,comment}

\begin{figure}
	\includegraphics[width=7cm]{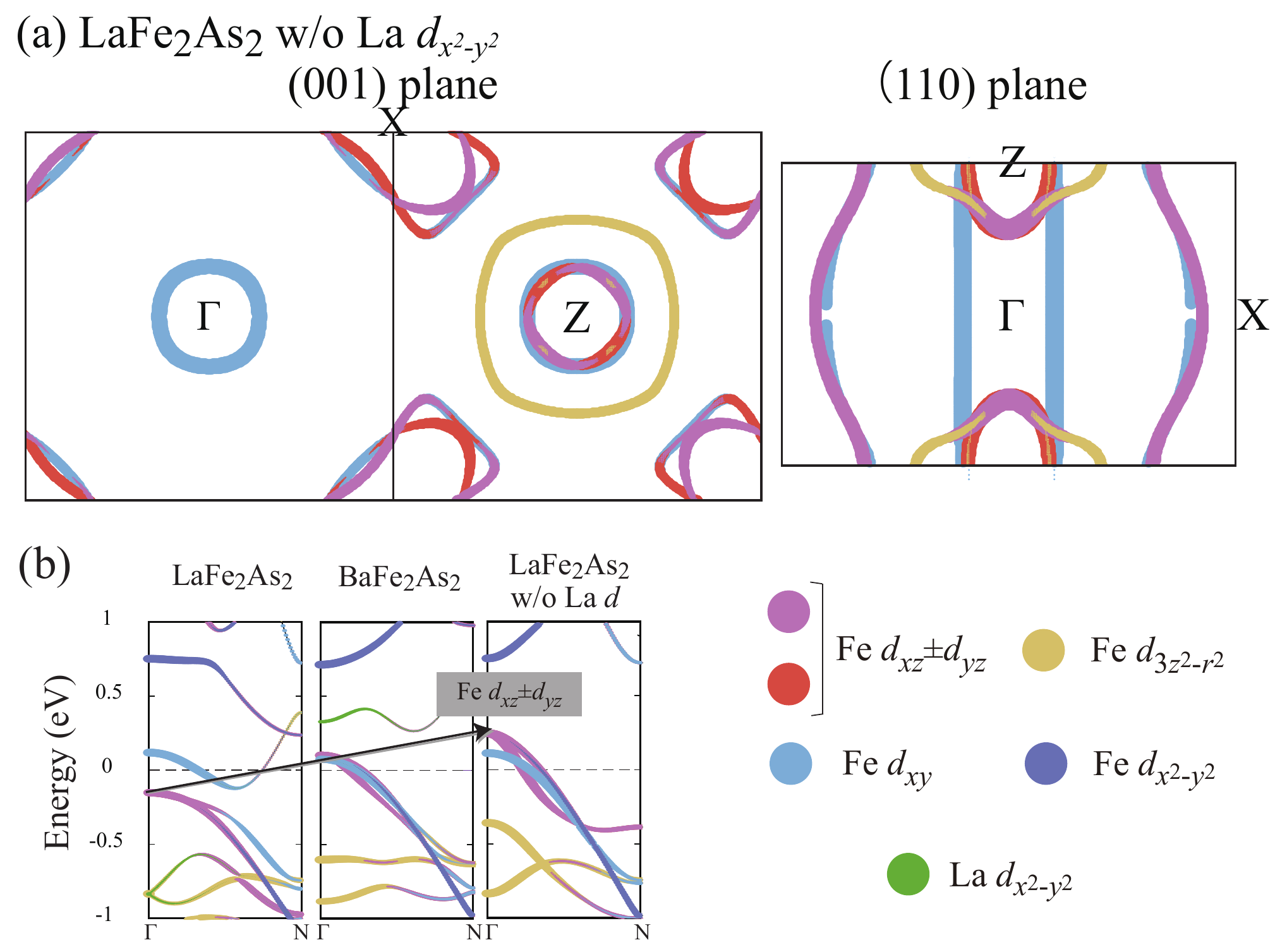}
	\caption{(a) The Fermi surface constructed from the 14-orbital tight binding model of uncollapsed LaFe$_2$As$_2$, 
	where the La $d_{x^2-y^2}$ orbital contribution is hypothetically removed.
	(b) The band structure of the 15-orbital models of uncollapsed LaFe$_2$As$_2$(left) and BaFe$_2$As$_2$(center), and the 10-orbital model of  
	uncollapsed LaFe$_2$As$_2$(right), where all the La $d$ orbitals are hypothetically removed from the 15-orbital model. 
	The Fermi energy of the hypothetical models of LaFe$_2$As$_2$ is fixed at that of the original LaFe$_2$As$_2$.
	}
	\label{fig3}
\end{figure}

Since the Fe $3d$ orbitals are expected to play a main role in the occurrence of superconductivity, 
it is interesting to look at the ``genuine Fe $3d$ appearance'' of the Fermi surface by hypothetically 
removing the La $d_{x^2-y^2}$ orbital contribution. For this purpose, we constructed a 14-orbital model, 
in which the hopping integrals among La $d_{x^2-y^2}$ and all the orbitals are neglected in the 15-orbital model. 
We adopt the same Fermi level as that of the 15-orbital model in order to 
directly compare the Fermi surfaces with and without considering the La $d_{x^2-y^2}$ orbital contribution.
The obtained Fermi surface is shown in Fig. \ref{fig3}(a).
The hole Fermi surfaces originating from the Fe $d_{xz/yz}$ orbitals are barely affected; 
they remain to be small pockets present only around the Z point, consistent with the heavy electron doping. 
However, most interestingly, there appears a cylindrical hole Fermi surface around the $\Gamma$-Z line, 
originating from the Fe $d_{xy}$ orbital, whose volume is similar to that in BaFe$_2$As$_2$. 
We will later come back to the origin of this apparent contradiction with the heavy electron doping. 
If we turn to the Fermi surfaces around the X point, there appear two cylindrical electron sheets, 
similar to those seen in BaFe$_2$As$_2$. 
Coming back once again to the original Fermi surface in Fig. \ref{fig2} from the above viewpoint, 
we can trace portions with strong Fe $d$ orbital weight to find that there essentially exist a $d_{xy}$ 
hole Fermi surface and a pair of electron Fermi surfaces usually seen in 122 iron-based superconductors. 
The bottom line here is that there actually exist electron and hole cylindrical Fermi surfaces 
originating from the $d_{xy}$ orbital, which can be favorable for spin-fluctuation-mediated superconductivity 
because this mechanism is based on the presence of electron and hole Fermi surfaces originating from the same orbital.

The robust survival of the $d_{xy}$ hole Fermi surface against heavy electron doping by elemental substitution in the blocking layer apparently resembles the case of the hydrogen-doped 
1111 systems,\cite{Hosono2011,Iimura2012,Iimura2017,Matsuishi2014}  
and is in contrast to the case of Co-doped BaFe$_2$As$_2$, where the elemental substitution takes place within the conducting layer, and only a rigid band shift of the Fermi level occurs as theoretically expected and experimentally observed.\cite{Fang2009,Liu2018}  
For the hydrogen-doped 1111, it was theoretically shown in Ref. \cite{Suzuki2014} that the substitution of oxygen with hydrogen leads to an increase of 
the positive charge in the blocking layer, which lowers the As $4p$ level (moves away from Fe $3d$) in the conducting layer.
This in turn reduces the indirect electron hopping between nearest neighbor Fe $3d_{xy}$ orbitals via As $4p$,
which raises the $d_{xy}$ hole band energy. 
One might expect a similar mechanism to work when Ba is substituted with  
La in BaFe$_2$As$_2$ because the positive charge in the blocking layer increases also in this case. 

\begin{table}[htb]
	\caption{The value of the on-site energy (unit: eV) measured from the Fermi energy for the 21-orbital models of uncollapsed LaFe$_2$As$_2$ and BaFe$_2$As$_2$.
    	$\Delta_{\rm La-Ba}$ is the energy difference between LaFe$_2$As$_2$ and BaFe$_2$As$_2$.
		\label{table1}}
		\begin{tabular}[t]{|c|c|c|c|c|} \hline
			\multicolumn{2}{|c|}{}         & La     & Ba     & $\Delta_{\rm La-Ba}$ \\ \hline
			            Fe & $d_{z^2}$     & -0.87 & -0.69 & -0.12 \\ \cline{2-5}
			               & $d_{xz/yz}$   & -0.55 & -0.43 & -0.12 \\ \cline{2-5}
			               & $d_{xy}$      & -0.61 & -0.49 & -0.12 \\ \cline{2-5}
			               & $d_{x^2-y^2}$ & -0.87 & -0.78 & -0.09 \\ \hline
			            As & $p_{x/y}$     & -1.99 & -1.71 & -0.28 \\ \cline{2-5}
						   & $p_{z}$       & -1.73 & -1.59 & -0.14 \\ \hline
					 La/Ba & $d_{z^2}$     &  1.97 &  2.96 & -0.99 \\ \cline{2-5}
			               & $d_{xz/yz}$   &  2.69 &  3.49 & -0.80 \\ \cline{2-5}
			               & $d_{xy}$      &  2.59 &  3.33 & -0.73 \\ \cline{2-5}
			               & $d_{x^2-y^2}$ &  1.83 &  2.74 & -0.92 \\ \hline
		\end{tabular}
\end{table} 

To see whether this is indeed the case, 
we construct 21-orbital models for both La and Ba cases, 
where not only Fe $d$ and La $d$ but also As $p$ orbitals are explicitly considered. As seen in Table \ref{table1},  
the on-site energies of both the As $p$ orbitals and the Fe $d$ orbitals are lowered by substituting Ba with La. In particular, 
the energy reduction of Fe $d$ orbitals and that of As $p_z$, through which the indirect Fe-Fe hopping mainly occurs, is nearly the same, 
which implies that the the robustness of the $d_{xy}$ Fermi surface is not due to the same mechanism as in hydrogen-doped 1111 systems.

Since the answer to the puzzle must lie in the difference between La and Ba, we go back to the 15-orbital model of LaFe$_2$As$_2$, 
and now hypothetically remove all the La $d$ orbitals to end up with a $15-5=10$ orbital model, whose band structure is shown in Fig. \ref{fig3}(b)(right).
Interestingly, we find that the position of the Fe $d_{xy}$ and $d_{xz/yz}$ hole bands at the $\Gamma$ point is reversed  
by totally removing the La $d$ orbitals, namely, the energy level of the $d_{xz/yz}$ hole bands becomes higher. This means that the energy of the Fe $d_{xz/yz}$ hole bands are suppressed due to the mixing of the La $d$ orbitals. What is surprising here is that the role played by the weakly hybridized La $d$ orbitals other than $d_{x^2-y^2}$ (note that there is small weight of other La $d$ orbitals near the Fermi level) is important because removing just the La $d_{x^2-y^2}$ orbital does not give rise to $d_{xz/yz}$ hole Fermi surfaces around the $\Gamma$ point as we have seen in the 14 orbital model (Fig. \ref{fig3}(a)). 
If we compare LaFe$_2$As$_2$ and BaFe$_2$As$_2$ in Fig. \ref{fig3}(b), 
the energy level of the $d_{xz/yz}$ hole bands of the latter is higher than that of the former. This is because the on-site energy of the Ba $d$ orbitals is higher than that of La $d$ (Table \ref{table1}). 
Actually, similar analysis for KFe$_2$As$_2$ shows that the 
energy of the $d_{xz/yz}$ band becomes even higher with respect to $d_{xy}$ (not shown), 
because the energy level of the K $d$ orbitals is higher than that of Ba $d$. 
Our conclusion here is that the $d_{xy}$ Fermi surface remains upon heavy electron doping in LaFe$_2$As$_2$ 
because the doped electrons selectively enter the $d_{xz/yz}$ bands, whose energy is significantly lowered by the La $5d$ hybridization. 
 
\begin{figure}
	\includegraphics[width=7cm]{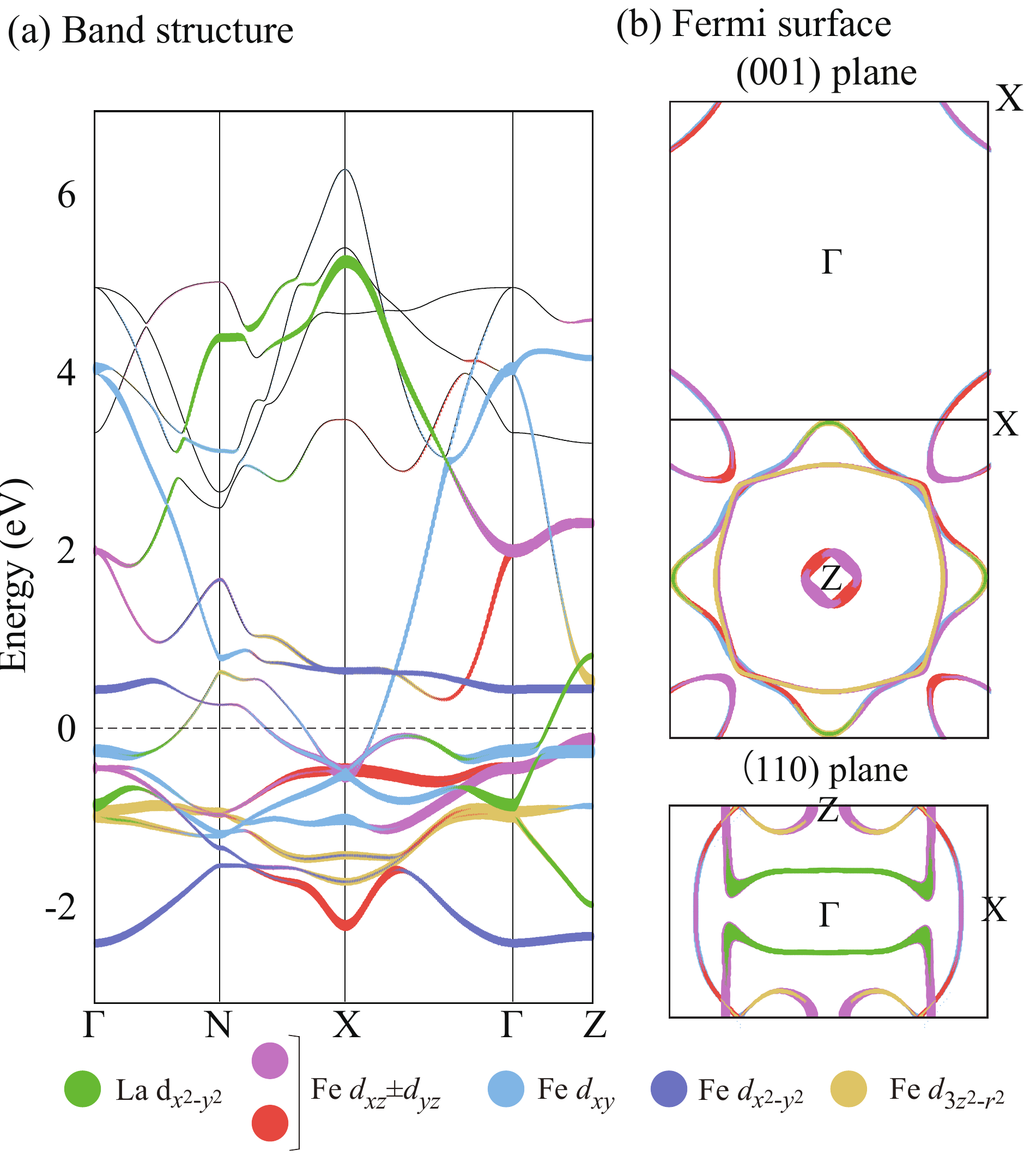}
	\caption{(a) The band structure and (b) the Fermi surface obtained from a 15-orbital tight binding model of collapsed LaFe$_2$As$_2$}
	\label{fig4}
\end{figure}

Next, we will discuss the difference of the band structure between 
the uncollapsed and collapsed phases.
Using the crystal structure of the collapsed phase described in Ref. \cite{Iyo2019}, 
we calculate the band structure and Fermi surface as shown in Fig. \ref{fig4}.
In the collapsed phase, the hole Fermi surface originating from the Fe $d$ orbitals almost disappears. 
In particular, the disappearance of the $d_{xy}$ hole Fermi surface, 
as compared to that in the uncollapsed phase, is due to the large Fe-As-Fe bond angle, namely, 
that of the uncollapsed phase is 110.8$^\circ$,
while that of the collapsed phase is 118.1$^\circ$. For such a large bond angle, 
the $d_{xy}$ bands sink deeply below the Fermi level in the iron based superconductors.\cite{Usui2012} 
From the stand point of spin-fluctuation-mediated pairing, 
the absence of $d_{xy/yz/xz}$ hole bands is unfavorable for superconductivity, 
so the absence of superconductivity in the collapsed phase can be naturally understood from this viewpoint.
We also note that the band structure of collapsed LaFe$_2$As$_2$ is similar to that of LaFe$_2$P$_2$, where superconductivity also does not emerge.\cite{Morsen1988,Razzoli2015,Blackburn2014} 

Finally, we comment on the superconducting gap structure of 
LaFe$_2$As$_2$ expected from the viewpoint of the orbital component on the Fermi surface. 
One may expect that the electron-electron interaction between the Fe $3d$ and La $5d$ orbitals are 
weak because they are spatially apart. 
Therefore, when the Fe $3d$ orbitals play a main role in the occurrence of superconductivity, 
one can expect that the superconducting gap is small (or vanishing) around the portion of the Fermi surface where 
the La $d$ orbital character is strongly dominating.
This may be one reason why $T_c$ is not as high as other 122 or 1111 compounds. 
Experimental probes are therefore expected to detect some kind of nodal feature in the gap structure. 

In summary, we calculated the electronic structure of uncollapsed and collapsed LaFe$_2$As$_2$. 
In both cases, the unique Fermi surface is a  consequence of the hybridization between the Fe $3d$ and La $5d$ orbitals.
The main difference between the uncollapsed and collapsed phases is the presence of the $d_{xy}$ hole Fermi surface in 
the former, which is clearly revealed by hypothetically removing the La $d_{x^2-y^2}$ orbital contribution. 
The robustness of the $d_{xy}$ Fermi surface against heavy electron doping is due to 
the decrease of the on-site energy of the La/Ba $d$ orbitals when Ba is replaced with La.
On the other hand, the Fe-As-Fe bond angle is too large in the collapsed phase, 
resulting in the disappearance of the $d_{xy}$ hole Fermi surface. 
The correspondence between the absence/presence of the $d_{xy}$ hole Fermi surface and 
absence/presence of $T_c$ can be naturally understood within the spin-fluctuation-mediated pairing scenario. 
Assuming this scenario, there may be room for further increasing $T_c$ by optimizing 
the volume of the $d_{xy}$ Fermi surface and/or removing the La $5d$ orbital contribution, by, say, 
partially substituting the La and/or As atoms.

We acknowledge A. Iyo and H. Mukuda for showing us the
experimental results for LaFe$_2$As$_2$ prior to publication, and 
motivating us to start the present study.
KK is supported by JSPS KAKENHI Grant Number JP18H01860.


\begin{thebibliography}{99}
\bibitem{comment}
In ref.[41], it was shown that LaFe$_2$P$_2$ and CaFe$_2$P$_2$ have nearly the same Fe $3d$ occupation, namely, 
replacing Ca by La does not result in electron doping into the Fe $d$ bands, but this does not apply to the relation between LaFe$_2$As$_2$ and BaFe$_2$As$_2$. 
Namely, although the hybridization with the Fe $3d$ bands is stronger for the La $d$ bands  than for the Ba $d$ bands, 
substituting Ba with La does result in electron doping, as can be seen from the comparison between Fig. \ref{fig2}(d) and Fig. \ref{fig3}(a), 
where the  $d_{xz/yz}$ hole Fermi surfaces significantly shrink and the electron Fermi surfaces bulge in LaFe$_2$As$_2$.
\end{thebibliography}
\end{document}